# Absence of large nanoscale electronic inhomogeneities in the Ba(Fe$_{1-x}$Co$_x$)$_2$As$_2$ pnictide

Y. Laplace[1], J. Bobroff[1*], F. Rullier-Albenque[2], D. Colson[2], and A. Forget[2]

[1]*Laboratoire de Physique des Solides, Univ. Paris-Sud, UMR 8502 CNRS, 91405 Orsay Cedex, France*
[2]*Service de Physique de l'Etat Condensé, Orme des Merisiers, CEA Saclay (CNRS URA 2464), 91191 Gif sur Yvette cedex, France*

23 july 2009

**Abstract.** $^{75}$As NMR and susceptiblity were measured in a Ba(Fe$_{1-x}$Co$_x$)$_2$As$_2$ single crystal for *x=6%* for various field *H* values and orientations. The sharpness of the superconducting and magnetic transitions demonstrates a homogeneity of the Co doping *x* better than *± 0.25%*. On the nanometer scale, the paramagnetic part of the NMR spectra is found very anisotropic and very narrow for *H//ab* which allows to rule out the interpretation of Ref.[6] in terms of strong Co induced electronic inhomogeneities. We propose that a distribution of hyperfine couplings and chemical shifts due to the Co effect on its nearest As explains the observed linewidths and relaxations. All these measurements show that Co substitution induces a very homogeneous electronic doping in BaFe$_2$As$_2$, from nano to micrometer lengthscales, on the contrary to the K doping.

**PACS.** 74.70.-b 74.62.Dh 76.60.-k 74.70.Dd

## 1 Introduction

Iron-based Pnictides display fascinating properties. Not only they show high temperature superconductivity, but they also have a very original phase diagram [1]. At zero doping, the Fe layers are semi-metals, with a commensurate spin density wave magnetic order. Upon carrier doping, the magnetism becomes incommensurate and is strongly reduced in amplitude. At high enough hole or electron doping, it turns into a superconductor with $T_C$ as high as *56K*, whose origin is not yet settled. True atomic scale coexistence between the magnetic and superconducting phases may even occur at intermediate doping range [2]. Hole or electron doping is achieved either by changing the composition of the intermediate layers (such as fluor at oxygen site in the LaOFeAs family, or K at Ba site in BaFe$_2$As$_2$ family), or by substitution at Fe site itself, for example by Co in BaFe$_2$As$_2$ [22]. Eventhough all families display similar phase diagrams, many of their properties strongly differ such as the value of $T_C$, the evolution of the magnetism versus doping, or even perhaps the symmetry of the order parameter [3]. It is thus essential to understand the actual role of the dopant on the physics of the Fe layers.

In this scope, one may expect that dopants not only change the carrier doping or the band structure, but could also affect the homogeneity of the Fe layer electronic state. Such situation has been widely discussed in high $T_C$ cuprate superconductors, where dopant induced disorder is advocated to explain many of the differences observed among cuprate families [4,5]. In Ba(Fe$_{1-x}$Co$_x$)$_2$As$_2$ pnictides, recent NMR [6] and tunneling [7] studies have argued that electronic properties are inhomogeneous on nanometer scale, possibly due to Co local effects. Inhomogeneity level was estimated from the NMR lineshape analysis to be so high that a Co *x=4%* sample would contain a sizeable amont of *x=8%* patches. If this were to be true, it would inturn have deep consequences on many properties of these componds, which should then be considered highly inhomogeneous. Situation seems even more inhomogeneous in (Ba$_{1-x}$K$_x$)Fe$_2$As$_2$ [8,9]. It is however surprising that such inhomogeneities do not reflect in macroscopic properties, such as the superconducting phase transitions or the Hall number change at the spin density wave transition which are both found very sharp [10].

In this study, we show for Ba(Fe$_{1-x}$Co$_x$)$_2$As$_2$ that a carefull analysis of the magnetic and superconducting transitions together with $^{75}$As NMR lineshape versus orientation of the applied field all demonstrate unambiguously that these componds are in fact much more homogeneous than previously argued. We propose that all NMR data including those of Ref.[6] can be accounted for by an anisotropic hyperfine coupling and/or chemical shift distribution due to Co local effect, but this has no impact on the electronic state of the Fe layers.

## 2 Experimental

Single crystals synthesis and characterization details are reported in [10] together with transport measurements which allowed to establish the phase diagram of fig.1-a which fits well other studies [22]. Most of this study was done on the sample studied in [2] at Co doping *x=6%*, which displays both superconductivity below $T_C=21K$ and frozen magnetism below $T_N=31K$ on local scale. The sample consists of a few single crystals aligned in a magnetic field in stycast epoxy along their c cristallographic axis, to get a better signal over noise ratio. This alignement was possible because the orbital component of the total susceptibility is anisotropic. We checked that one single crystal gives the exact same linewidth and spectral shape than the sample of aligned crystals, both in perpendicular and parallel direction with the applied field. Orientation of the sample was performed *in-situ* taking advantage of the high sensitivity of the NMR line position upon the field orientation.

[75]As NMR was performed in a fixed applied magnetic field ranging from *3* to *7.5 Tesla*, using standard pulse echo sequences and Fourier Transform recombinations. Short enough delays between pulses (ranging from *10* to *50 μsec*) were used to avoid any effect of the transverse $T_2$ relaxation. Longitudinal $T_1$ relaxation measurements were recorded after a saturation of the signal using a $\pi/2$ pulse. Macroscopic susceptibility was measured in a MPMS-SQUID magnetometer in a zero-field cooled procedure, taking into account the demagnetizing factors associated with its thin and rectangular shape [2].

## 3 Mesoscopic homogeneity

Both the superconducting and magnetic phase transitions widths allow to constrain the doping distribution in our sample. These transitions are indeed broadened by any inhomogeneous Co doping if it occurs on a lengthscale large enough for the transition temperature to be defined. This kind of inhomogeneity will thus be referred as "mesoscopic inhomogeneity" i.e. on lengthscales larger than a few nanometers. The superconducting transition is monitored by the Meissner effect measured by the magnetization decrease below $T_C$ (fig.2b). The magnetic transition is monitored by the intensity wipeout of the NMR central line (fig.2c). Below $T_N$, the spectrum strongly broadens as shown in [2]; the total intensity does not change, but the intensity measured only over a central narrow frequency range decreases sharply at $T_N$. But due to technical reasons, this T-dependence of the central line intensity is a more accurate criterium of the magnetic transition than that of the full width of the spectrum.

In the case of such a doping distribution $P(x)$, we can deduce its effect on both transitions and comparison with experiment will give a typical doping distribution upper bound. To get such estimate, we use a simple model where the proportion $P(x)$ of the sample at an effective Co doping $x$ becomes superconducting and perfectly diamagnetic below $T_C(x)$ and magnetically ordered below $T_N(x)$. This should result in a decrease of the total macroscopic susceptibility at $T<T_C(x)$ proportional to $P(x)$ and a drop of the NMR intensity at $T<T_N(x)$ also proportional to $P(x)$. Indeed, when the Fe moments freeze below $T_N$, the NMR intensity of the central line is observed to wipeout [9] because of large broadening and relaxation effects.

We consider for P(x) three gaussian doping distributions centered at *x=6%* of full widths at half maximum (FWHM) *0.25%*, *0.5%*, and *1%* as displayed in fig1-a. Using $T_N(x)$ and $T_C(x)$ of fig.1-a we can deduce the corresponding expected superconducting and magnetic phase transitions for each distribution as plotted in fig.1-b and 1-c respectively. These simulations are compared to the experiment on our *x=6%* sample. The experimental superconducting transition width is compatible with a distribution of doping with FWHM between 0.5% and 1%.

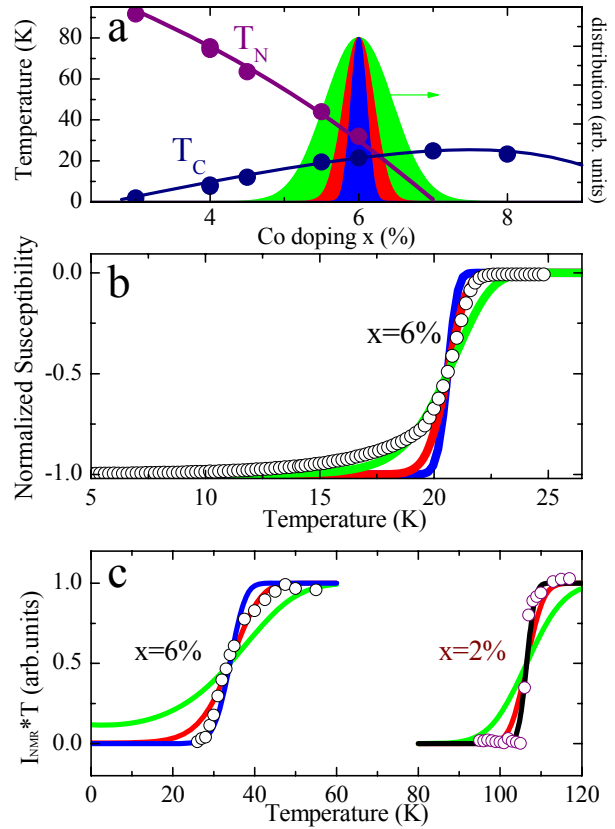

**Fig. 1.** Phase diagram of Ba(Fe$_{1-x}$Co$_x$)$_2$As$_2$ from Ref. [10] together with the three modeled Gaussian doping distributions of FWHM *0.25%*, *0.5%* and *1%* (blue, red and green respectively) b) simulated superconducting transition for the three distributions (solid curves with the same color code), and experimental data (white circles) for the

*x=6% sample c) simulated magnetic transitions (solid curves) and experimental data (white circles) for x=6% and x=2%*

However, this is only an upper bond since other effects can broaden the superconducting transition, such as non perfect screening when London depth is comparabable to the size of the samples, which may occur close to $T_C$ [11].Using a similar analysis, the magnetic transition is well described with a distribution of FWHM between *0.25%* and *0.5%*, which is compatible with the upper bound given by the susceptibility. These estimates of the mesoscopic doping distribution made free of any adjustable parameter are remarquably low : at *x=6%*, the true doping is ranging only from typically 5.8% to 6.2% and Co doping can then be considered highly homogeneous, at least in regards to the bulk properties. We also report the NMR intensity data for a *x=2%* sample which magnetically orders at $T_N=106$ *K*. Hereagain, the sharpness of the transition points toward a very homogeneous doping distribution of maximum FWHM *0.25%*.

We stress that this analysis does not state that the shapes of the transitions are only accounted for by Co doping distribution. But if even a low doping distribution is to occur at mesoscopic scale, it should lead to much broader transitions than the one observed experimentally.

## 4 Nanoscopic homogeneity

### 4.1 The $^{75}$As NMR lineshape

The analysis of the bulk transitions does not allow to exclude electronic inhomogeneities if they vary on very short distances. Indeed, a variation of the doping content at the nanometer scale could be averaged out and result in sharp superconducting and magnetic transitions but still affect locally the electronic properties such as local susceptibility. $^{75}$As NMR is well suited to address this possible "patchy" situation, since each $^{75}$As nucleus is coupled mostly only to its four near neighbour Fe through short range hyperfine couplings. Any variation of the electronic properties even on nanometer range will result in a distribution of the $^{75}$As NMR spectrum and dynamics. For example, in the case of high Tc cuprates, $^{63}$Cu Nuclear Quadrupole Resonance allowed to show the existence of large inhomogeneities in La$_{2-x}$Sr$_x$CuO$_4$[12] while $^{17}$O and $^{89}$Y NMR allowed to demonstrate high homogeneity in the case of YBa$_2$Cu$_3$O$_{6+y}$ [13].

The central line of the NMR spectrum for the *x=6%* sample is plotted on fig.2 for two fields orientations ($H_0//c$ axis and $H_0//ab$) at *T=90K*. Due to the axial symmetry of the paramagnetic phase, the quadrupolar parameters along a and b are equal, and the quadrupolar asymmetry is *η=0*. The $^{75}$As NMR frequency of the central line can then be written for an angle *θ* between the applied field $H_0$ and *c* axis :

$$\nu(\theta) = \bar{\gamma}H_0 + \bar{\gamma}H_0(K_{ab}\sin^2\theta + K_c\cos^2\theta)$$
$$- \frac{\nu_{Qc}^2}{2\bar{\gamma}H_0}\left[-\frac{27}{8}\cos^4\theta + \frac{30}{8}\cos^2\theta - \frac{3}{8}\right]$$

where $\bar{\gamma}$ is the $^{75}$As gyromagnetic ratio and $\nu_{Qc}$ is the c axis quadrupolar frequency.

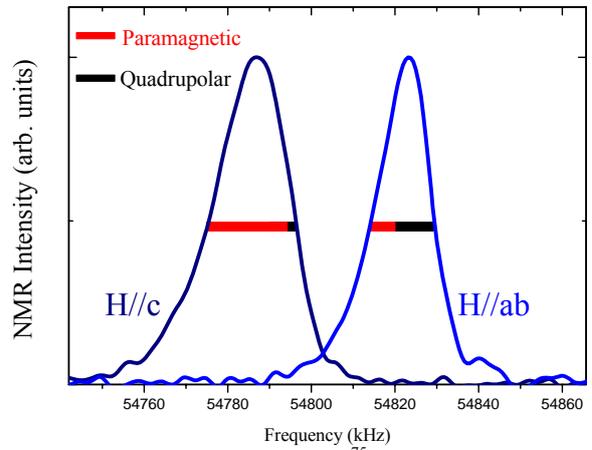

**Fig. 2.** Central line of the $^{75}$As NMR spectrum in Ba(Fe$_{94\%}$Co$_{6\%}$)$_2$As$_2$ at *T=90K* for field *H//c* and *H//ab* with *H=7.5T*. Respective paramagnetic (red) and quadrupolar (black) contributions deduced from the field dependence are found very anisotropic.

The paramagnetic shifts *K* are proportional to the local electronic susceptibility through:

$$K = K_{spin} + K_{chem} = A_{hf}\chi_{spin}/\mu_B + K_{chem}$$

where $A_{hf}$ are the spin hyperfine couplings between $^{75}$As nucleus and Fe electrons and $K_{chem}$ is the sum of all orbital contributions, both from valence and inner shell electrons. In the c and ab directions of $H_0$, $\nu(\theta)$ simplifies to:

$$\nu(H//c) = \bar{\gamma}H_0(1+K_c)$$
$$\nu(H//ab) = \bar{\gamma}H_0(1+K_{ab}) + \frac{3}{16}\frac{\nu_{Qc}^2}{\bar{\gamma}H_0}$$

Both the paramagnetic and quadrupolar shifts may contribute to the width Δν of the NMR line if they are distributed for any reason. We note hereafter these paramagnetic and quadrupolar contributions $\Delta M$ and $\Delta Q$. These can be determined using their different field $H_0$ dependences, as $\Delta M \propto H_0$ while $\Delta Q \propto 1/H_0$. If the two effects simply convolute, we get:

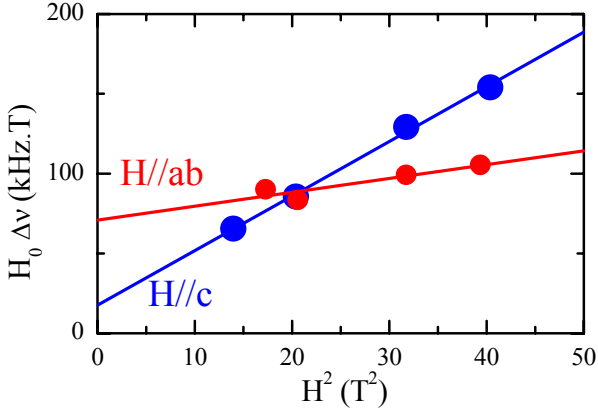

**Fig. 3.** field dependence of $H_0 \Delta \nu$ (where $\Delta \nu$ is the width of the NMR central line) at $T=50K$ for the two field orientations in Ba(Fe$_{94\%}$Co$_{6\%}$)$_2$As$_2$. The slope, proportional to the paramagnetic contribution, is smaller along ab than along c.

$$H_0 \Delta \nu = H_0 (\Delta Q + \Delta M) = a + bH_0^2$$

where $\Delta Q = a/H_0$ and $\Delta M = bH_0$. In fig.3, $H_0 \Delta \nu$ plotted for each field orientation is indeed found to be linear in $H_0^2$, which confirms the validity of the above description. We deduce from linear fits that at $H_0=7.5T$ and $T=90K$:

$$\Delta M_{//c} = 19.1 \pm 2 \text{ kHz and } \Delta Q_{//c} = 1.7 \pm 1 \text{ kHz}$$
$$\Delta M_{//ab} = 6 \pm 1 \text{ kHz and } \Delta Q_{//ab} = 9 \pm 1 \text{ kHz}$$

These contributions are represented by red (paramagnetic) and black (quadrupolar) bars at half maximum on the spectra in fig.2 (no account should be given to the location of the arrows over the spectrum: the two broadenings are mixed within the line and convolute each other). At $H_0=7.5T$, the broadening of the NMR line along c axis is almost fully paramagnetic, whereas it is more quadrupolar along ab. This is due to the fact that no quadrupolar shift is expected in $\nu(H//c)$ while both effects contribute to $\nu(H//ab)$.

The paramagnetic broadening $\Delta M_{//c} = 19.1 kHz$ of the NMR line along c axis is similar to that measured by Ning et al. [6]. In Ref.[6], the authors interpreted this broadening to be due to the spatial distribution of the electronic spin susceptibility originating from nanoscale electronic inhomogeneities induced by Co doping. *If this was to be true, a very similar paramagnetic broadening should be observed for H//ab. On the contrary, our present results clearly demonstrate that the broadening along ab is much smaller* [16]. This argument can be made more quantitative. The spin susceptibility $\chi_{spin}$ is almost isotropic in Ba(Co$_{1-x}$Fe$_x$)$_2$As$_2$ [14] so any doping distribution should distribute $\chi_{spin}$ similarly for H//c and H//ab. The spin hyperfine coupling anisotropy is $A_{hf}^c / A_{hf}^{ab} \approx 0.7$ at $x=0$ [15] and not expected to be much doping dependent at small Co dopings. Along the proposal of Ref.[6], we thus expect the paramagnetic width for H//ab to be about 27kHz, i.e. about 4 times larger than our experimental determination of the paramagnetic contribution. We therefore conclude that *the present data rules out any large doping distribution*, on the contrary to Ning's conclusions.

## 4.2 The $^{75}$As NMR relaxation rate $1/T_1$

Relaxation measurements were also argued to demonstrate a large doping distribution because the longitudinal relaxation rate $1/T_1$ was found to vary by about *30 %* along the NMR spectrum [6, 9]. Our own measurement of $1/T_1$ at various positions of our spectra shows a similar but slightly smaller distribution (of about *20%*). Such small distribution can originate from various mixed effects detailed hereafter. But the fact that we observe a similar and even larger distribution for $1/T_1$ for H//ab while paramagnetic broadening is much smaller here again shows that this distribution does not necessarily comes from a doping distribution.

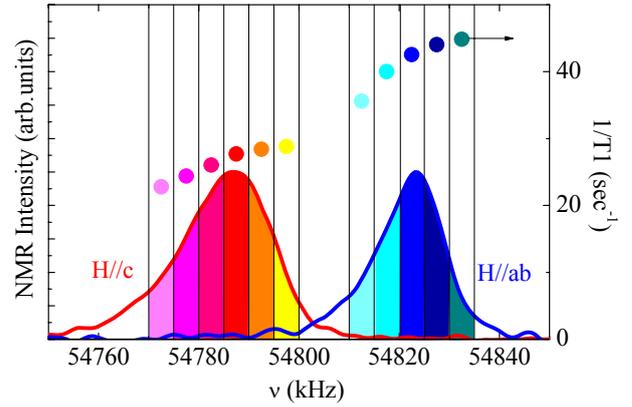

**Fig. 4.** frequency dependence of the inverse of the spin lattice relaxation time $1/T1$ over the NMR line for the two field orientations at $T=50K$ and $H=7.5T$ in Ba(Fe$_{94\%}$Co$_{6\%}$)$_2$As$_2$.

## 4.3 Possible sources of NMR broadenings

Let us stress again our two main results: the paramagnetic NMR broadening is very anisotropic (about 3 times smaller along ab than along c) and its magnitude along ab is very small, since the paramagnetic width of the NMR line $\Delta M_{//ab}$ is more than 20 times smaller than its shift. One source of broadening which could explain these findings is a distribution of the hyperfine coupling itself, while its average value would remain constant. Any Co substituted at Fe site is likely to modify the orbital environment of its near and next nearest neighbour As atoms, hence their hyperfine couplings $A_{hf}$. The first neighbors might not be detected here because they are too shifted, as observed in the strongly overdoped composition [21]. But a smaller effect on further neighbors could be expected. This would result in a distribution of the NMR frequency with the same type of paramagnetic broadening as that of an electronic doping distribution. It could also explain the variations of $1/T_1$ since $1/T_1 \propto A_{hf}^2$. Furthermore, it leads to a broadening which increases with Co content, as also observed in [6] In order to explain the observed anisotropy of $\Delta M$, the effect of Co on hyperfine paths could be more effective along c axis than along ab axis. But an accurate theoretical description of the orbitals involved and how they could be affected by Co-Fe substitution is needed to elaborate more from present data. In the same manner, Co doping could also affect the chemical-orbital shift $K_{chem}$ of $^{75}$As, here again leading to an additional broadening. This effect could be sizeable here since $K_{chem}$ is anomalously large (of the order of 0.2%).

Another source of broadening could be due to RKKY-like effects around each Co. Indeed, in metals or correlated systems, atomic substitution is known to induce RKKY-like decaying alternated oscillations of the local paramagnetic susceptibility in its viscinity [17]. Such oscillations of $\chi_{spin}$ lead to a broadening of the NMR line following a 1/T Curie-like dependence. This behavior has been observed in the presence of Sn impurities in Ba$_{1-x}$K$_x$Fe$_2$As$_2$ crystals [18]. In our case, a similar but much smaller 1/T contribution to the linewidth is also observed which suggests a small impurity contribution. However, this effect should not display sizeable anisotropy since electronic susceptibility is isotropic here. Testing how the anisotropies and linewidths evolves with Co doping and among other pnictides family should help understand the respective importance of these various sources of broadenings.

## 5 Discussion and conclusion

In our *x*=6% sample, the quantitative analysis of the bulk superconducting and magnetic transitions shows that the actual doping distribution is smaller than about *±0.25%* on mesoscopic scale. This ensures that the nominal Cobalt contents used in the phase diagrams reported so far can be trusted. On a nanoscopic scale, our NMR study of the field-dependence further demonstrates the absence of any large nanoscale inhomogeneity of the electronic properties. This seems to contradict Scanning Tunnelling Microscopy measurements which showed that the superconducting gap magnitude varies by about *±20%* on the surface of superconducting *x=7%* and *10%* samples [7]. This would correspond to much larger doping distributions than found in our study. However the superconducting gap is known to be very sensitive to any source of disorder [4]. The Co-Fe substitution could then partly act as a local impurity-like disorder as suggested by the RKKY discussion of the NMR data, which would account for the STM results with no need for doping inhomogeneity to occur. We also stress that these are surface measurements, while all NMR or macroscopic susceptibility measure the volume of the samples.

Our results are to be contrasted with the K-doped BaFe$_2$As$_2$ where NMR and μSR measurements demonstrate a more inhomogeneous situation [8, 9, 19, 20]. The magnetic transition monitored by NMR is much wider, and in the intermediate regime, the K-doped compound is even found to segregate on mesoscopic scale between superconducting and magnetic domains. Furthermore, the NMR width is found not only much larger but also more dependent on temperature in K-doped than in Co-doped sample [9]. The observed 1/T type of dependence could signal that the disorder induced RKKY-like effects are more pronounced for K than for Co. This is very surprising since one would expect in-plane Co substitution to have more detrimental impurity-like effects than out-of plane K defects, as in high Tc cuprates. However, here, the more metallic nature of the Fe planes and the smaller electron correlations may make Fe-pnictides less sensitive to in-plane substitutions than high T$_C$ cuprates. In contrast, non isovalent K substitution at Ba site could locally distort the crystal structure and modify the distances and angles between As and Fe orbitals, which are known to be decisive in the physics of the pnictides.

Understanding how these various types of dopants affect the local and macroscopic properties of the pnictides will be a key ingredient in the understanding of the pnictides and their phase diagram, especially when comparing the various families.

*Note added in proof*: after the completion of this manuscript, we became aware of a new study by Ning et al. [21] on a highly overdoped Ba(Fe$_{74\%}$Co$_{26\%}$)$_2$As$_2$ which shows that the NMR shift *K* on $^{75}$As near neighbors of Co is the same as that far from Co. This corroborates our picture of a homogeneous electronic susceptibility in the Fe layers.

*We acknowledge for very fruitfull discussions H. Alloul, F. Bert, F. Bouquet, V. Brouet, Y. Ihara, M.H. Julien, G. Lang, H. Luetkens, and P. Mendels.*

* Electronic address: bobroff@lps.u-psud.fr